**Programmable photonics enabled by ferroionic two-dimensional materials**

Ghada Dushaq [1*], Srinivasa R Tamalampudiand[1], Solomon Serunjogi[1], Mahmoud Rasras [1,2*]

*[1]Photonic Research Lab (PRL) Engineering Division, Department of Electrical and Computer Engineering, New York University Abu Dhabi, PO Box 129188, Abu Dhabi, UAE*
*[2]NYU Tandon School of Engineering, New York University, New York, New York USA*

## Abstract

Artificial intelligence (AI) models are scaling rapidly, exposing fundamental limitations in conventional computing architectures, where the physical separation of memory and computation imposes substantial energy and latency overheads. Optical neural networks offer a viable solution by enabling high-throughput, low-latency computation through the intrinsic parallelism of light. However, the scalability of photonic computing is constrained by the lack of materials that can provide low-loss, energy-efficient, and nonvolatile control of optical phase. Here, we discuss the emerging role of ferroionic two-dimensional materials as a platform for programmable photonics. Unlike conventional ferroelectrics, where polarization arises from bounded lattice distortions, ferroionic systems enable field-driven ionic redistribution, providing access to a continuum of stable and reconfigurable states. We outline how these material dynamics can be exploited to realize multi-level photonic states and nonvolatile phase control. Furthermore, we highlight a back-end integration strategy that introduces multifunctionality, including tunable optical nonlinearities, into pre-fabricated photonic circuits without compromising optical performance. By bridging material-level dynamics with device- and system-level functionality, ferroionic materials provide a pathway toward reconfigurable, low-loss, and energy-efficient photonic computing architectures, opening new opportunities for neuromorphic and adaptive photonic systems.

## Overview

The accelerating advancement of artificial intelligence (AI) models is exposing inherent limitations in traditional computing architectures[1–3]. The intrinsic separation of memory and computation in the von Neumann architecture introduces significant inefficiencies due to data movement[4]. The exponential growth of model sizes has shifted the performance bottleneck toward data movement, making inter-unit data transfer the critical constraint on system efficiency[5]. Copper interconnects carrying data between thousands of cores dissipate energy at every hop. On the other hand, biological neural systems, rely on tightly coupled units where memory and computation happen in the same physical location achieving remarkable computational power at orders of magnitude lower energy consumption[6,7]. The path forward lies in platforms where computation and storage are intrinsically integrated[8,9]. Photonic integrated circuits (PIC), which perform the core operations of neural network inference directly in the optical domain, offer precisely this opportunity, provided the right materials are found to make them programmable[10–14].

## Optical computing as an alternative framework

Optical neural networks (ONNs) provide a viable solution to address these challenges[15–17]. Photons propagate without resistive loss, enabling light-based computation to bypass the fundamental data-movement bottleneck that constrains electronic systems.[14,18] This offers orders-of-magnitude improvements in bandwidth, latency, and energy efficiency by performing linear algebra in the analog optical domain[13,14,19]. In addition, the ability to multiplex information across multiple wavelengths enables parallel data processing within the same physical channel which supports massive parallelization essential for complex AI models[20–22]. A key advantage of optical systems lies in their ability to perform matrix-vector multiplication (MVM) through interference and superposition at the speed of light. These operations, which form the computational core of neural networks (~90% of compute resources are devoted to MVM), can be realized directly within photonic circuits without sequential processing steps[13,23–26]. System-level models indicate that the energy advantage of optical processors over electronic accelerators grows with model size, suggesting a scalable rather than diminishing return. As a result, photonic computing provides the real-time processing speeds needed for high-throughput AI workloads while maintaining a superior energy profile[27–30].

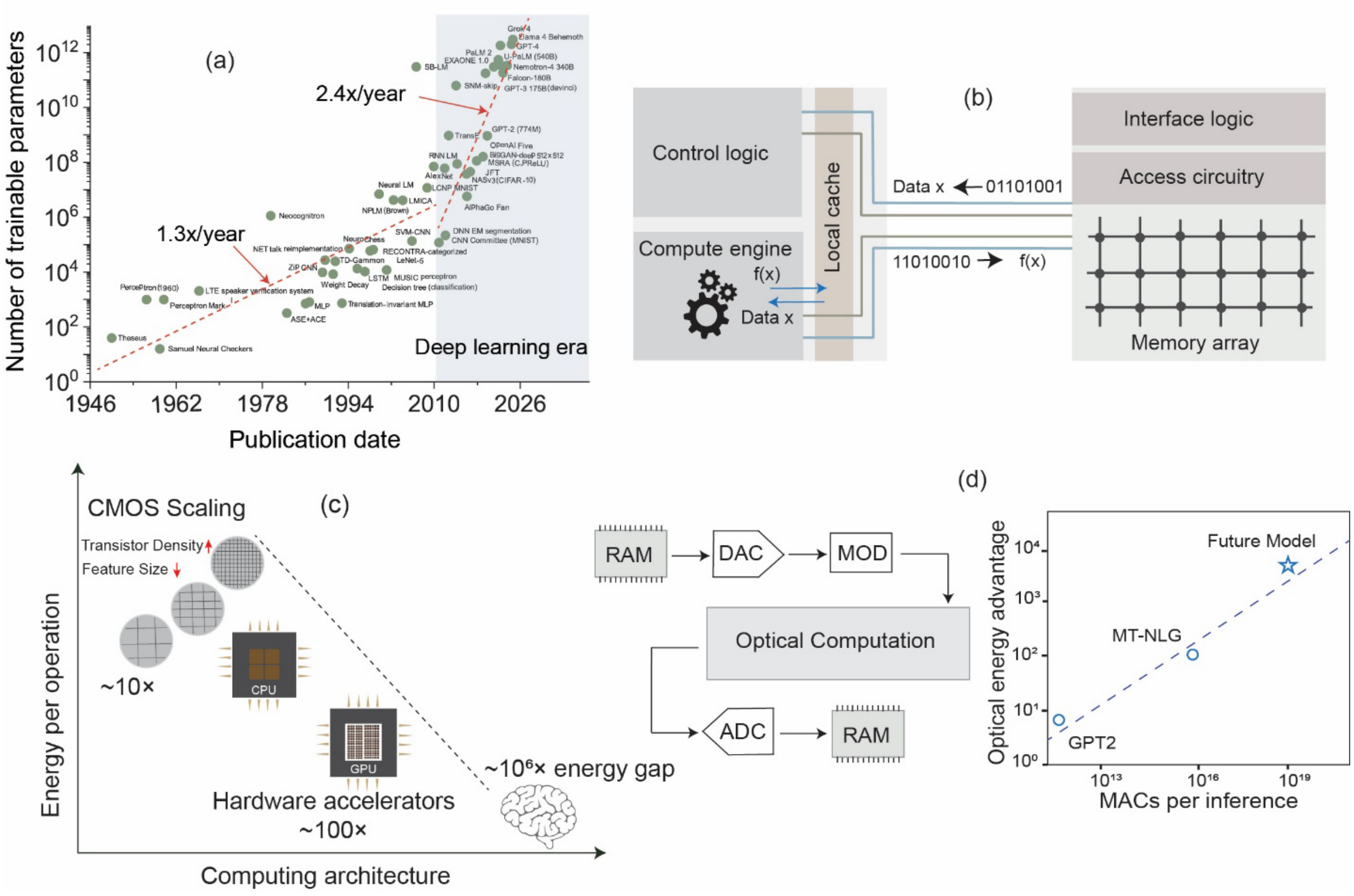

Figure 1| **Scaling limits of electronic computing and the emergence of photonic accelerators.** (a) Growth in the number of trainable parameters across frontiers AI models illustrates the accelerating computational demand that motivates hardware innovation (data from the Epoch AI model dataset, available: https://epoch.ai/data/ai-models). (b) The physical separation of computation and memory units in conventional von Neumann designs

necessitates constant data flow between processing units and memory arrays. The substantial delay and energy overhead caused by this data transport (the so-called memory wall) is facilitated by copper interconnects. (c) CMOS scaling and specialized hardware accelerators such as GPUs and ASICs have enhanced computational efficiency, yet these improvements are still insignificant in comparison to the orders of magnitude of the energy efficiency observed in biological neural systems. (d) Photonic computing performs matrix–vector multiplication directly in the optical domain through interference and superposition, bypassing resistive data movement. The energy advantage of optical processors grows with model size, according to system-level models, indicating a scalable approach to get around electronic design energy limitation. (the energy consumption data for optical accelerators adapted from optical transformers[29,30])

## The materials bottleneck

Despite the advantages offered by ONNs, the scalability of photonic computing systems is hindered by the lack of suitable tunable materials[31,32]. In current photonic integrated circuits (PICs), phase shifters are essential components that impose a controllable phase delay on a guided optical mode to encode a computational weight[13,33]. Traditional silicon modulation mechanisms rely on techniques such as the thermo-optic effect, which is constrained by high static power consumption (often in the milliwatt range per tuner) and considerable cross talk[34,35]. Similarly, free-carrier dispersion in silicon suffers from doping-induced optical losses[36]. Crucially, these mechanisms are volatile and cannot maintain a given state without continuous power, leading to Joule heating and undermining energy-efficient computing[37]. These limitations highlight a fundamental challenge in achieving compact, low-loss, and energy-efficient control of optical phase in integrated photonic circuits. To align with future CMOS-compatible roadmaps, the industry must transition toward hybrid material platforms[38]. In such approaches, optical phase shifters rely on the guiding properties of silicon or SiN platforms, while the active, nonvolatile, multi-level low-loss, and energy-efficient tunability is derived from the electro-optic properties of the cladding.[39,40] The hybrid integration of promising materials like phase-change materials (PCM) and thin-film ferroelectrics is now the focus of substantial industry and academic efforts[41,42]. For instance, nonvolatile optical phase control based on PCMs such as $Ge_2Sb_2Se_4Te$ (GSST) has been demonstrated[43]. These materials switch between amorphous and crystalline phases with a large refractive index contrast. However, PCM-based devices perform at most a handful of discrete states, require electrical or optical pulses to heat the material above its crystallization or melting temperature. The cyclic thermal load of repeated amorphization and crystallization introduces mechanical and chemical inhomogeneity that fundamentally limits device endurance, a challenge that requires complex nanostructuring strategies to mitigate[44]. Additionally, systems using GST suffer from residual absorption in the crystalline phase that accumulates with device count[45]. Thin-film ferroelectric such as lithium niobate offers ultrafast Pockels-effect modulation with ultra-low propagation loss, but it is not CMOS-native and requires wafer bonding[46,47]. The ideal platform would combine the nonvolatility of phase-change materials, the low optical loss of passive silicon photonics, analog programmability over many distinguishable levels, and monolithic compatibility with back-end semiconductor processing.

## Materials as the engine of technological shifts

Two-dimensional (2D) materials have emerged as promising candidates due to their atomically thin structure and compatibility with back-end integration. Their ability to maintain strong electronic and optical performance at reduced dimensions makes them ideal for next-generation optoelectronic devices[48,49]. From an industrial standpoint, the development of new technologies is directly related to identifying system-level bottlenecks and introducing materials that can overcome these limitations, hence positioning 2D materials as a viable material platform for future semiconductor systems[38,50].

Nevertheless, the majority of 2D material-based phase shifters remain limited by either volatility or insufficient modulation efficiency[49,51]. Recently, significant attention has been given to 2D multiferroic materials such as $CuCrP_2S_6$ (CCPS) and $CuInP_2S_6$ (CIPS)[52,53]. Optically, these materials possess a moderate bandgap ranging from 1.4 to 3.5 eV, making them transparent in the short-wave infrared (SWIR) range[54]. Additionally, their high refractive index, with relatively small contrast to silicon, leads to low optical loss and improved mode matching[54,55]. In contrast, conventional 3D semiconductors with high refractive index tend to be absorptive in the telecom band, highlighting the advantage of 2D multiferroics in offering both transparency and high-index in the SWIR regime. From an electrical perspective, these materials are classified as type-II multiferroics, exhibiting multiple ferroic orders and notable ionic conductivity that couples spin, valley, and electric dipole degrees of freedom[56,57]. By directly coupling material dynamics to optical response, these systems offer the potential to realize reconfigurable photonic devices.

## Multi-level photonic states via material dynamics

Conventional ferroelectrics rely on collective localized ionic displacements within rigid crystal lattice, where the switching of polarization is governed by domains nucleation and domain-wall movement[58]. Although, these processes can exhibit complex dynamics, they are constrained by the bounded nature of lattice distortions leading to limited accessible configurational space. As a result, achieving highly reconfigurable and continuously programmable functionalities remains challenging. In addition, ferroelectric thin films often suffer from polarization degradation driven by depolarization fields and size dependent effects which hinder device scaling progress[59].

In contrast, Van der Waals (vdW) ferro ionic crystals such as CIPS operate differently and represent a paradigm shift by coupling ferroelectricity with ionic conductivity[60]. These materials exhibit liquid-like solid dynamics where their polarization is carried by highly mobile copper ions that occupy multiple symmetry-inequivalent sites and can migrate both within a single vdW layer (intralayer, in-plane) and across the interlayer gap (interlayer, out-of-plane)[61]. Because polarization is determined by the spatial distribution of these ions rather than by a collective lattice mode, the system supports a continuum of intermediate states between the two polarization extremes. This configurational richness absent in conventional ferroelectrics and inaccessible to phase-change materials is the physical basis for analog-optical programmability.

Identifying and controlling ion hopping pathways at the nanoscale is essential for hardware reliability and predictable device behavior. For instance, in CIPS copper ions has multiple allowed sites and they can migrate through two primary pathways: intralayer and interlayer hopping. By designing the voltage pulse amplitude, duration, width and pulse-count, a configurable kinetics for multi-level states can be achieved[45,62]. Tuning the voltage pulse parameters leads to different programmability response of CIPS and they are classified in three types based on the spatial distribution of copper ions. Type I dominated by intralayer migration where copper ions displaced within a single layer and this mode is triggered by using low voltage, fast and short-duration pulses. An intermediate mode is characterized by coupled kinetics and represented as Type II where intralayer and interlayer switching are both triggered using higher voltage amplitude or longer pulse duration compared to Type I. In Type III, the dominant mechanism is the interlayer migration where copper ions hop across vdW gap.  This is high energy barrier process and requires high voltage amplitude or significantly longer pulse width. As a result, a single CIPS device can operate as a fast-volatile weight updater, a gradual synaptic accumulator, or a stable nonvolatile memory element simply by modulating the electrical pulse waveform[56,62]. This programmable state space extends beyond conventional binary switching, enabling continuous analog encoding essential for photonic neural computation.

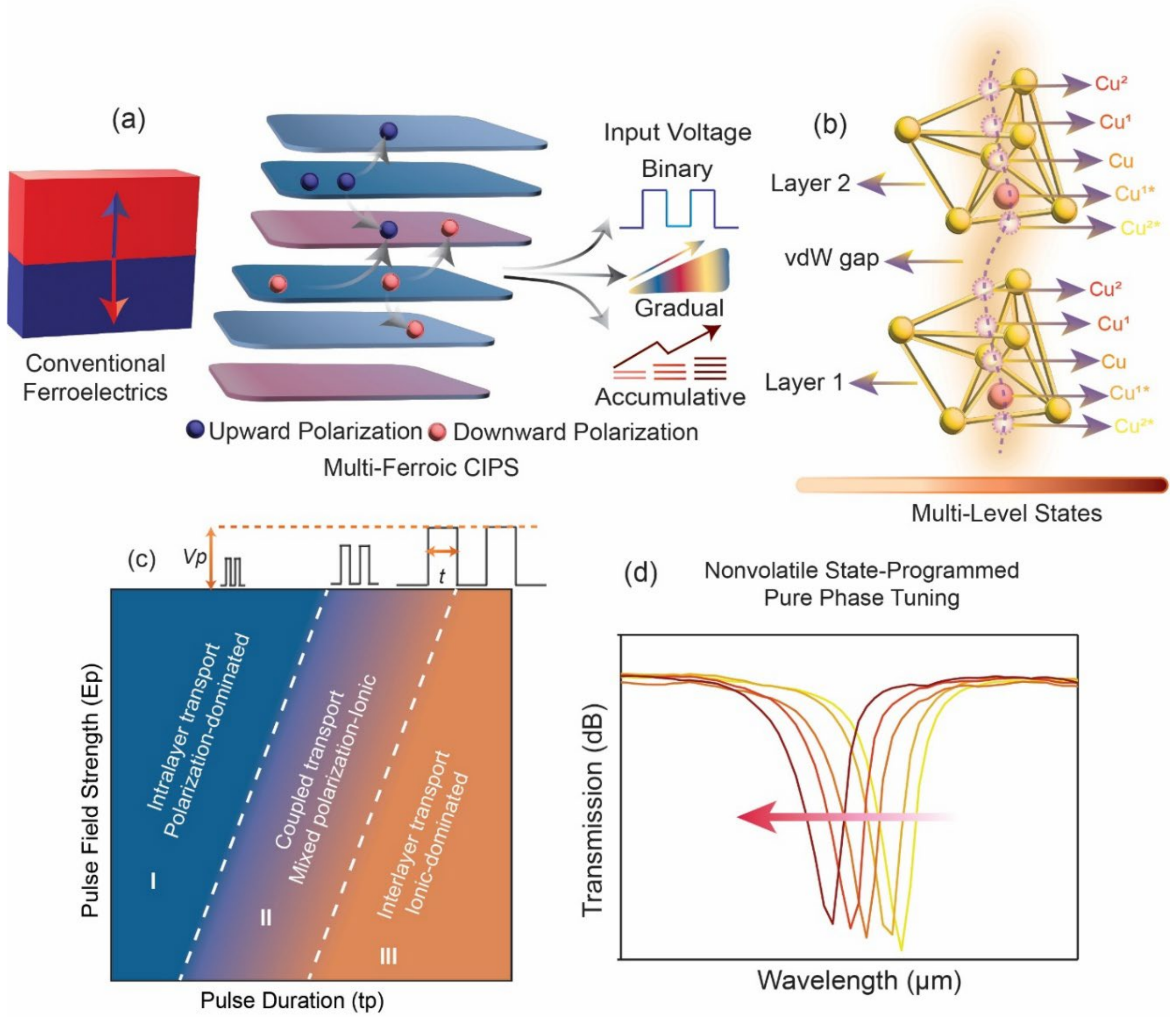

Figure 2| **Multi-level photonic state programmability via ionic dynamics in CIPS.** (a) Schematic comparison of polarization switching in a conventional ferroelectric versus ferroionic CIPS. Conventional ferroelectrics exhibit binary polarization with states governed by bounded lattice distortions, CIPS supports both discrete and continuous responses, enabling binary, gradual, and accumulative state evolution under electrical stimuli. (b) Atomic-scale depiction of copper-ion redistribution within and across van der Waals layers, illustrating intralayer (in-plane) and interlayer (out-of-plane) hopping pathways that collectively define the accessible configurational space. (c) Phase diagram of pulse-driven switching dynamics as a function of pulse amplitude (Vp) and duration (tp), illustrating the transition from intralayer polarization-dominated response (I), through coupled ionic–polarization behavior (II), to interlayer ionic transport-dominated regime (III). (d) Experimental demonstration of nonvolatile, state-programmed pure phase tuning in an integrated silicon microring resonator, where successive ionic configurations induce deterministic resonance shifts without introducing optical loss, enabling multi-level photonic functionality[54].

## Back-end integration for multifunctional photonic systems

Scalable photonic computing is constrained not only by the availability of suitable materials but by their integration without degrading established platforms. Back-end integration offers a practical pathway, enabling functional materials to be incorporated onto pre-fabricated photonic circuits while preserving the optical guiding layer. In this scheme, phase modulation is introduced through the cladding rather than the core, allowing pure phase control without added optical loss or disruption of front-end processes, and enabling low-loss, energy-efficient reconfigurability.

This approach is gaining traction, particularly in systems leveraging phase-change materials and active photonics, including recent industrial efforts (e.g., IMEC)[63,64]. Within this framework, multiferroic 2D materials provide an expanded functional space. By utilizing a dry-transfer stamping process in which vdW flakes or continuous films are mechanically exfoliated and placed onto pre-fabricated photonic structures at room temperature, with no wet chemistry or elevated thermal budget. Because the active material modulates the cladding rather than the waveguide core, the optical mode is perturbed only evanescently, and the incremental absorption introduced by the 2D material is negligible at SWIR wavelengths where CIPS and CCPS are intrinsically transparent[54,55]. The net result is that nonvolatile phase tuning, multi-level, and electrically controlled can be introduced onto a foundry-fabricated chip without degrading its baseline insertion loss or requiring any modification to the front-end process flow.

This back-end compatibility has broader significance. It means that ferroionic integration is not in competition with mature silicon photonic platforms but is additive to them. Thus, a chip designed and fabricated for passive routing or sensing can subsequently receive active, nonvolatile programmability through a post-processing stamp. The approach is therefore aligned with the emerging paradigm of chiplet heterogeneous integration, in which distinct functional layers are optimized independently and combined at the packaging stage.

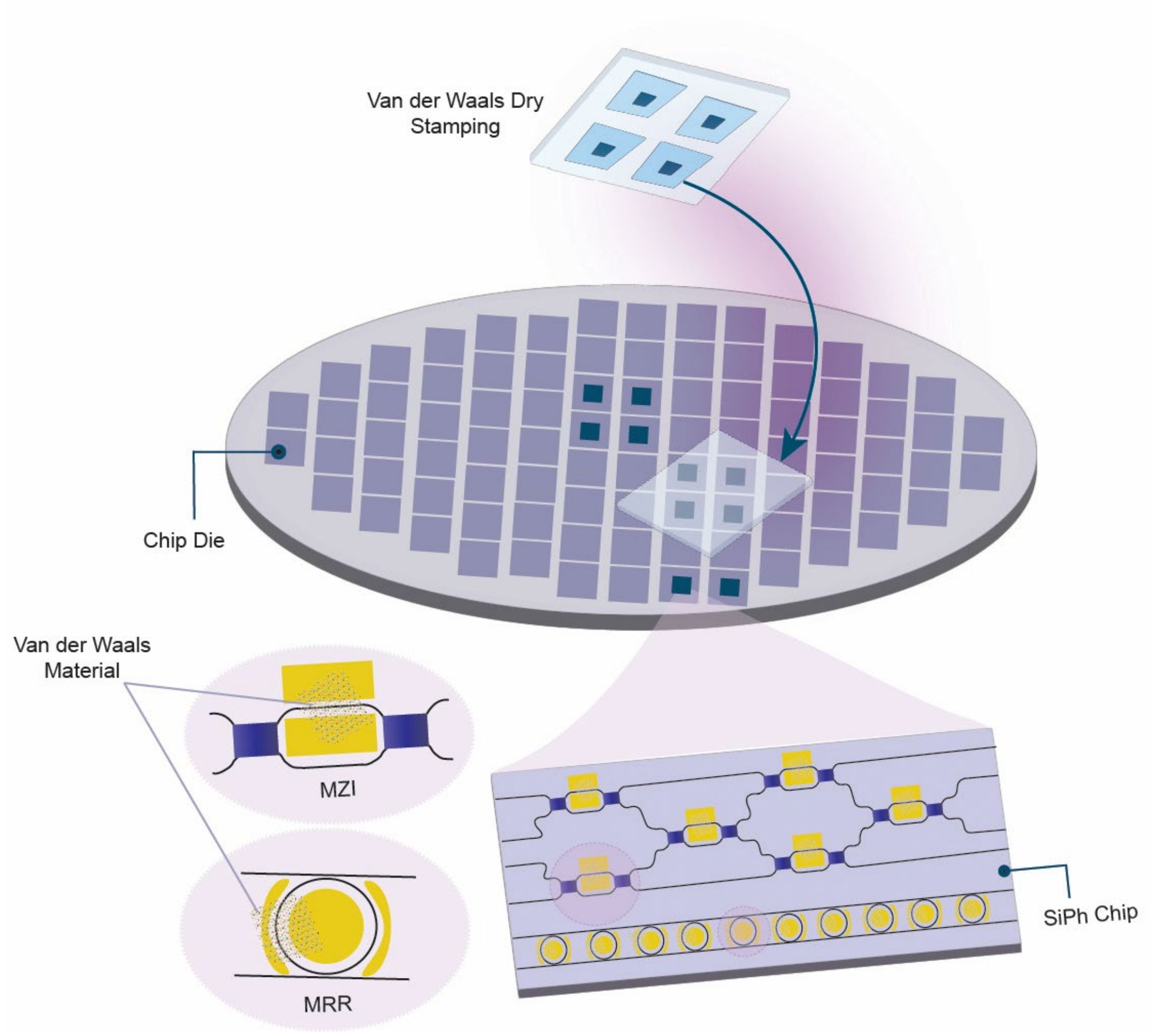

Figure 3| **Material integration for programmable photonic devices.** Schematic illustrating the transfer and selective placement of van der Waals ferroionic materials onto pre-fabricated silicon photonic chips using a dry stamping process. The material is positioned over key photonic elements, including Mach–Zehnder interferometers and micro-ring resonators, enabling localized modification of the optical response through the cladding without altering the underlying waveguide structure. This integration approach introduces reconfigurable and nonvolatile functionality while preserving the integrity and scalability of the photonic platform.

## Beyond phase modulation

A photonic neural network requires two functional primitives: linear weighted summation (implemented through interference in a programmable coupler or Mach–Zehnder mesh) and a nonlinear activation function (analogous to the rectifier or sigmoid in a digital network)[65]. In electronic implementations these operations are straightforwardly separated. In integrated photonics, the nonlinear activation has historically been the challenge. Silicon is centrosymmetric (vanishing $\chi^2$) and its third-order nonlinearity ($\chi^3$) is modest, requiring long interaction lengths or high optical powers that are incompatible with the low-loss, compact devices needed for large-scale integration.

Ferroionic 2D materials offer a direct route to co-integrated optical nonlinearity[66]. CIPS and CCPS are non-centrosymmetric in their polar phase and exhibit finite second-order susceptibility ($\chi^2$) that is entirely absent in centrosymmetric silicon at SWIR wavelengths[53,67,68]. Direct measurements of second-harmonic generation confirm strong $\chi^2$ responses in both materials[68]. This alone provides a route to on-chip parametric processes frequency conversion, entangled photon pair generation, optical rectification that silicon photonics cannot support natively. Whether the $\chi^3$ response at SWIR telecom wavelengths is comparably enhanced remains to be quantified, but the polar symmetry and high refractive index of these materials suggest a promising nonlinear platform which worth systematic characterization.

When integrated as cladding layers on microring resonators, the resonant enhancement of the circulating optical field amplifies the effective nonlinear interaction by orders of magnitude relative to a straight waveguide. Electrically tuning the ionic configuration of the ferroionic layer modifies the resonator transfer characteristic it's through-port transmission as a function of wavelength and thereby continuously reshapes the effective nonlinear transfer function experienced by the optical signal. This means that a single back-end-integrated ferroionic resonator could enable simultaneous realization of programmable phase element (through the linear refractive index change) and as a tunable nonlinear activation (through the electrically reconfigurable $\chi^2/\chi^3$ response). Linear weighted summation and nonlinear activation are co-localized within the same physical device a capability that has no precedent in silicon photonics with a single material.

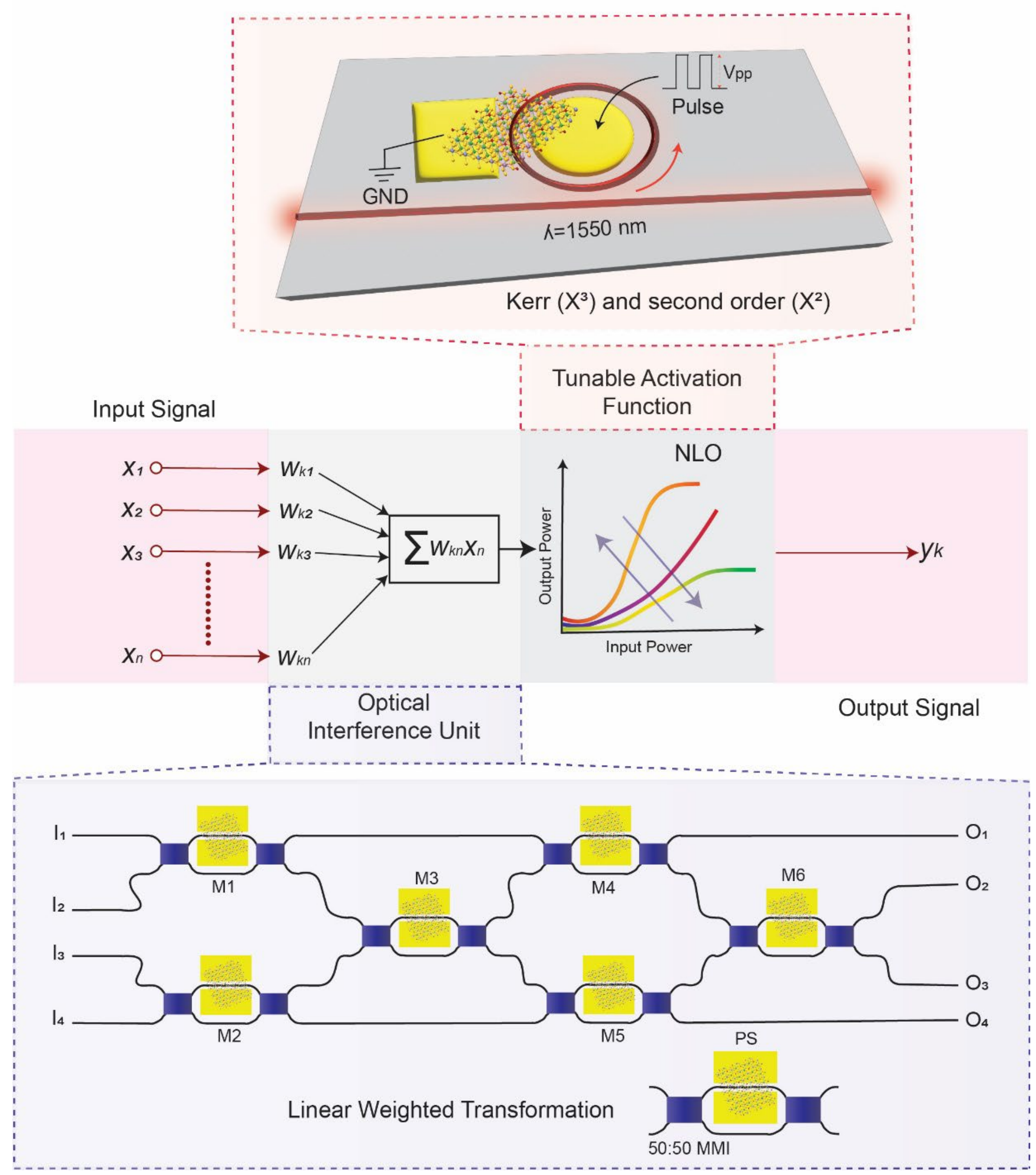

Figure 4 | **Back-end integrated photonic for tunable activation and multifunctional computing.** (Top) Schematic of a back-end integrated resonator in which a ferroionic 2D material is placed over a silicon microring, enabling electrically programmable phase modulation through the cladding without perturbing the underlying waveguide. The material simultaneously supports second- and third-order nonlinear optical responses ($\chi^2$ and $\chi^3$), providing access to electrically tunable nonlinear transfer characteristics. (Middle) Conceptual framework of a photonic neural unit, where input optical signals are linearly combined through interference to perform weighted summation, followed by a tunable nonlinear activation function enabled by the material response, yielding the output signal, all within the optical domain and without optoelectronic conversion. (Bottom) Example implementation of a linear optical interference unit based on cascaded multimode interferometers and phase shifters, illustrating how passive photonic circuits perform matrix–vector multiplication, while nonlinearity is introduced locally through the

back-end integrated material. Together, this architecture enables compact, low-loss, and reconfigurable photonic computing by co-integrating linear and nonlinear functionalities on fully optical neural networks.

## Outlook and future directions

The integration of ferroionic materials into photonic platforms opens a promising pathway toward programmable and multifunctional optical systems. However, several fundamental and technological challenges remain. A central question is how to achieve precise and reproducible control of ionic dynamics at the nanoscale, where copper-ion migration is influenced by local defect density, grain boundaries, and interface chemistry. These factors can vary between individual flakes and across transferred films, leading to stochastic intermediate switching states and, consequently, weight-encoding errors in photonic neural networks. Developing growth and transfer protocols that produce uniform ionic landscapes, whether through approaches such as chemical vapor deposition of large-area CIPS films, controlled defect engineering, and electrode geometries that guide ion migration deterministically, remains a key materials challenge.

A second challenge concerns the trade-off between switching speed and nonvolatility. Long-range interlayer ionic transport, which enables highly stable nonvolatile states, is intrinsically slower than electronic or lattice-mediated switching mechanisms. For inference at the edge, where weights are updated infrequently and fast optical readout is the primary requirement, this trade-off is acceptable. For online learning applications that require rapid, continuous weight updates, engineering thinner active layers with lower migration barriers or exploiting the faster Type I intralayer switching regimes will be necessary. At the same time, the long-term endurance and cycling stability of such systems must be carefully evaluated, particularly in the context of large-scale integration.

At the system level, the co-integration of linear and nonlinear functionalities enabled by back-end material integration presents new opportunities for photonic computing architectures. The ability to realize nonvolatile weights alongside tunable activation functions within the same physical platform could enable compact and fully integrated optical neural networks. Extending this approach to large-scale photonic integrated circuits will require advances in fabrication uniformity, CMOS-compatible integration, and circuit-level design methodologies that explicitly account for coupled ionic and photonic dynamics.

Looking ahead, the convergence of ferroionic material physics with integrated photonics offers a framework in which memory, linear transformations, and nonlinearity can be co-localized within a unified platform. While still at an early stage, this approach provides a compelling pathway toward reconfigurable, adaptive, and energy-efficient photonic systems, with potential relevance for next-generation AI and neuromorphic hardware.

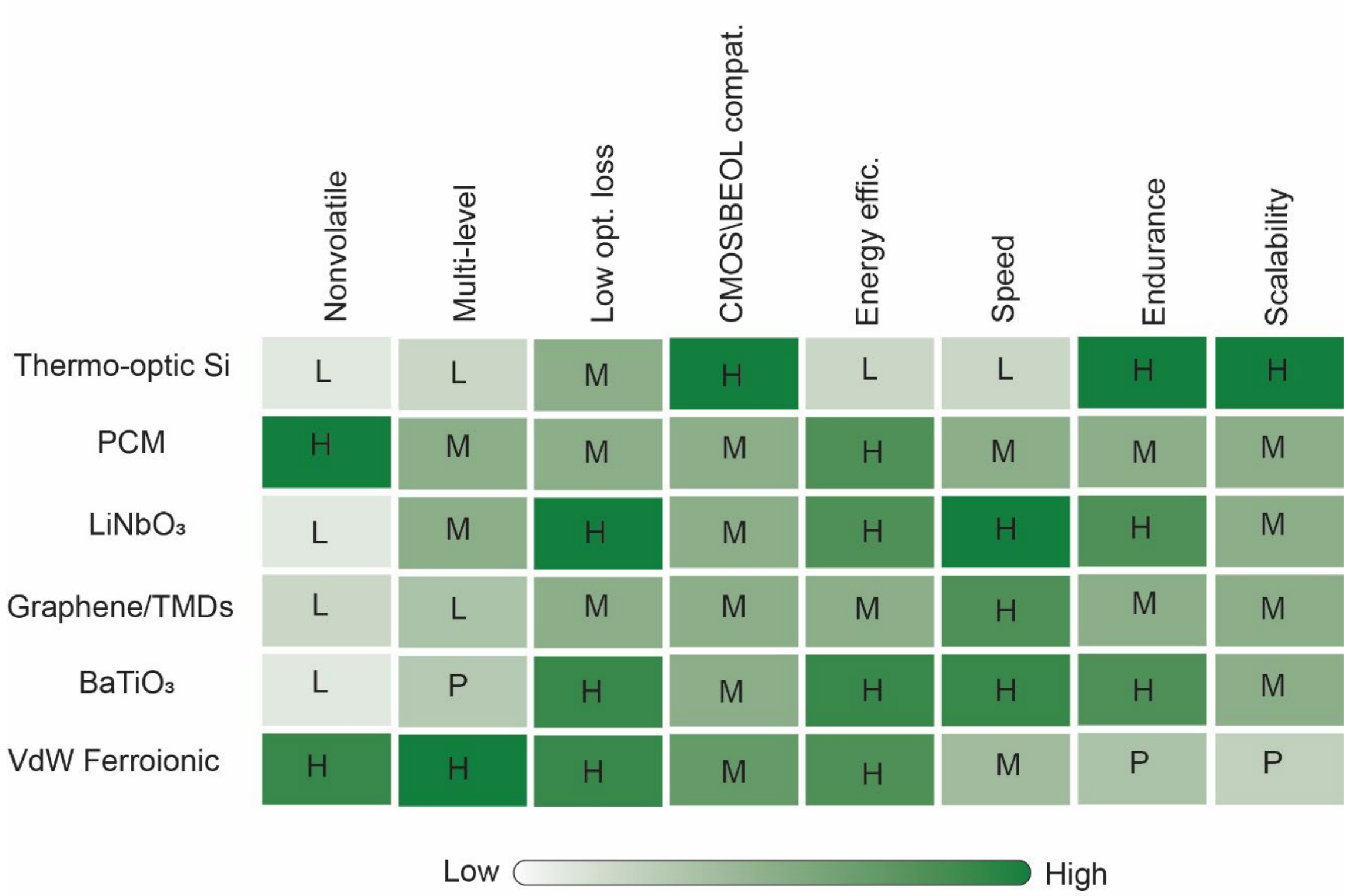

Figure 5| **Benchmarking of material platforms for integrated photonics phase-shifters**. Qualitative comparison of representative photonic modulation platforms across key metrics relevant to programmable and energy-efficient photonic computing, including nonvolatility, multi-level programmability, optical loss, CMOS/back-end-of-line (BEOL) compatibility, energy efficiency, speed, endurance, and scalability. Ratings are shown qualitatively as high (H), medium (M), partial (P), or low (L), based on reported device characteristics and integration considerations.

**Acknowledgements**

The authors acknowledge support from New York University Abu Dhabi and NYUAD Photonics and Core Technology Platform Facility (CTP).

**Competing interests**

The authors declare no competing interests.